\documentclass[pra,twocolumn,showpacs,preprintnumbers,amsmath,amssymb]{revtex4-1}
\makeatletter
\def\@dotsep{4.5}
\makeatother
\usepackage[dvips]{graphicx}

\usepackage{amsmath}
\usepackage{amsbsy}
\usepackage{color}
\setcitestyle{square}
\usepackage{epstopdf}
\DeclareGraphicsExtensions{.pdf,.eps,.png,.jpg,.mps}

\definecolor{textcolor}{cmyk}{0,0,0,1}
\definecolor{magenta}{rgb}{1,0,1}
\definecolor{green}{rgb}{0,1,0}
\definecolor{red}{rgb}{1,0,0}

\begin{document}

\title{
Topologically confined states at corrugations of gated bilayer graphene}
\author{M. Pelc}
\affiliation{Institute of Physics, Faculty of Physics, Astronomy and Informatics, Nicolaus Copernicus University, Grudziadzka 5, 87-100 Toru\'n, Poland}
\author{W. Jask\'olski}
\affiliation{Institute of Physics, Faculty of Physics, Astronomy and Informatics, Nicolaus Copernicus University, Grudziadzka 5, 87-100 Toru\'n, Poland}
\author{A. Ayuela}
\affiliation{Centro de F\'isica de Materiales, CFM-MPC CSIC-UPV/EHU, Donostia International Physics Center (DIPC)}
\affiliation{Departamento de F\'isica de Materiales, Facultad de Qu\'imicas, UPV-EHU, 20018 San Sebasti\'an, Spain}
\author{Leonor Chico}
\affiliation{Instituto de Ciencia de Materiales de Madrid (ICMM), Consejo Superior de Investigaciones Cient\'ificas (CSIC), C/ Sor Juana In\'es de la Cruz 3, 28049 Madrid, Spain}

\date{\today}

\begin{abstract}
We investigate the electronic and transport properties of gated bilayer graphene with one corrugated layer, which results in a stacking AB/BA boundary. When a gate voltage is applied to one layer, topologically protected gap states appear at the corrugation, which reveal as robust transport channels along the stacking boundary. With increasing size of the corrugation, more localized, quantum-well-like
states emerge. These finite-size states are also conductive along the fold, but in contrast to the stacking boundary states, which are gapless, they present a gap. We have also studied periodic corrugations in bilayer graphene; our findings show that such corrugations between AB- and BA-stacked regions behave as conducting channels that can be easily identified by their shape. 

\end{abstract}


\maketitle


\section{\label{sec:intro} Introduction}
The distinctive electronic properties of monolayer graphene can be modified by stacking more graphene layers on top \cite{McCann_2006,Guinea_2006}. For instance, Bernal (also called AB-stacked) bilayer graphene (BLG) shows a parabolic dispersion relation at low energies, and a gap can be opened by the external electric field applied perpendicular to the system \cite{Ohta_2006, Castro_2010}. Although the experimentally obtained energy gaps are moderate, such a gap opening is impossible in monolayer graphene or bilayer graphene with other stackings, so it is of great importance for making electronic devices based on graphene. Interestingly, experimental results indicate that the transport gap in Bernal-stacked BLG is smaller than the optical one \cite{Oostinga_2008,Zhang_2009,Szafranek_2010}. There is a substantial dispersion of the gap values, depending on whether the sample is suspended or not, and in the latter case, the type of substrate employed. In fact, even for zero bias some samples showed a transport gap \cite{Bao_2012, Velasco_2012,  Freitag_2012, Freitag_2013}, while others presented a metallic behavior \cite{Feldman_2009, Weitz_2010,Maher_2013}. Many-body effects have been invoked to explain such differences \cite{Bao_2012,Stroucken_2013}; however, another plausible explanation has arisen recently, namely, the existence of stacking boundaries in bilayer graphene.

In monolayer graphene, domain walls can be atomically sharp, consisting on grain boundaries made of topological defects with associated edge states \cite{Lahiri_2010, Yu_2011, Huang_2011}. These grain boundaries have been extensively studied in multilayer graphene \cite{Ayuela_2014,Vancso_2013} and even in carbon nanotubes \cite{Jaskolski_2005}. In BLG, boundaries composed of topological defects also occur \cite{Alden_2013, Zhang_2013_JAP, Brown_2012}, but domain walls can as well consist of a stacking dislocation between AB and BA regions (equivalent to the  so-called AB-AC boundaries). Such structures have been experimentally identified by various techniques, evidencing their ubiquity in 
bilayer graphene \cite{Lin_2013, Alden_2013, Hibino_2009, Ju_2015}. In contrast to boundaries made of topological defects, stacking boundaries are not atomically sharp. They can appear as strained regions, which produce a gradual transition between AB and BA-stacked graphene. The connection between the two stacking regions can take place in different forms: not only by tensile or shear strain, but also by corrugations of one of the layers \cite{Katsnelson_2008,Koshino_2013, Vaezi_2013,San_Jose_2014}. The reversal of stackings has significant influence on the electronic transport, it lowers the conductance between the AB and BA regions and leads to the appearance of the topologically protected states localized in the transition region.

Indeed, gapped bilayer graphene has been recently predicted to be a quantum valley Hall insulator \cite{Martin_2008,Zhang_2013,Vaezi_2013}. It can have chiral edge states associated to the two valleys propagating in opposite directions, which are topologically protected if valley mixing is precluded. These robust states were predicted to appear in electric-field domain walls \cite{Martin_2008}, where two regions differ in the sign of the applied voltage; however, these systems are not easy to implement experimentally. In fact, they are equivalent to an AB/BA stacking boundary, as those produced in strained  bilayer graphene with a uniform applied voltage, which also show these robust gapless states  \cite{Jung_2011,Koshino_2013,Vaezi_2013,San_Jose_2014}. Stacking boundaries have been experimentally observed \cite{Alden_2013, Lin_2013, San_Jose_2014} and there is recent evidence of the existence of topologically protected states in them \cite{Ju_2015}.

Ripples or corrugations in single layer graphene were theoretically predicted to appear as a consequence of thermal fluctuations \cite{Fasolino_2007}, and have been observed with transmission electron microscopy \cite{Meyer_2007, Meyer_2007_Nat}. These ripples can lead to interesting electronic \cite{Guinea_2008, Yu_2010, Santos_2012}, magnetic \cite{Ayuela_2012}, and chemical properties \cite{Boukhvalov_2009}, mostly due to the associated charge inhomogeneity and the changes in hybridization. In fact, folding graphene has been put forward as a way to modify its properties  \cite{Cranford_2009,Costa_2013,Chen_2014};  also, folded ribbons have been proposed as graphene-based electronic connectors between edges or graphene layers \cite{Gonzalez_2012}. All these schemes are pointing to a novel path to tune the characteristics of graphene-based systems, giving rise to the so-called "origami" graphene  \cite{Cranford_2009,Costa_2013,Hallam_2015}.

Folds, corrugations and wrinkles can also appear on top of multilayer graphene \cite{Meyer_2007,Zhu_2012}. This type of extended defect may lead to a stacking boundary between the two sides of the fold. Contrary to boundaries produced by shear and tensile strains, the transport properties of corrugated bilayer graphene with a stacking boundary have not been yet investigated so thoroughly as in the strained bilayer.

\begin{figure}[thpb]
\centering
\includegraphics[width=\columnwidth]{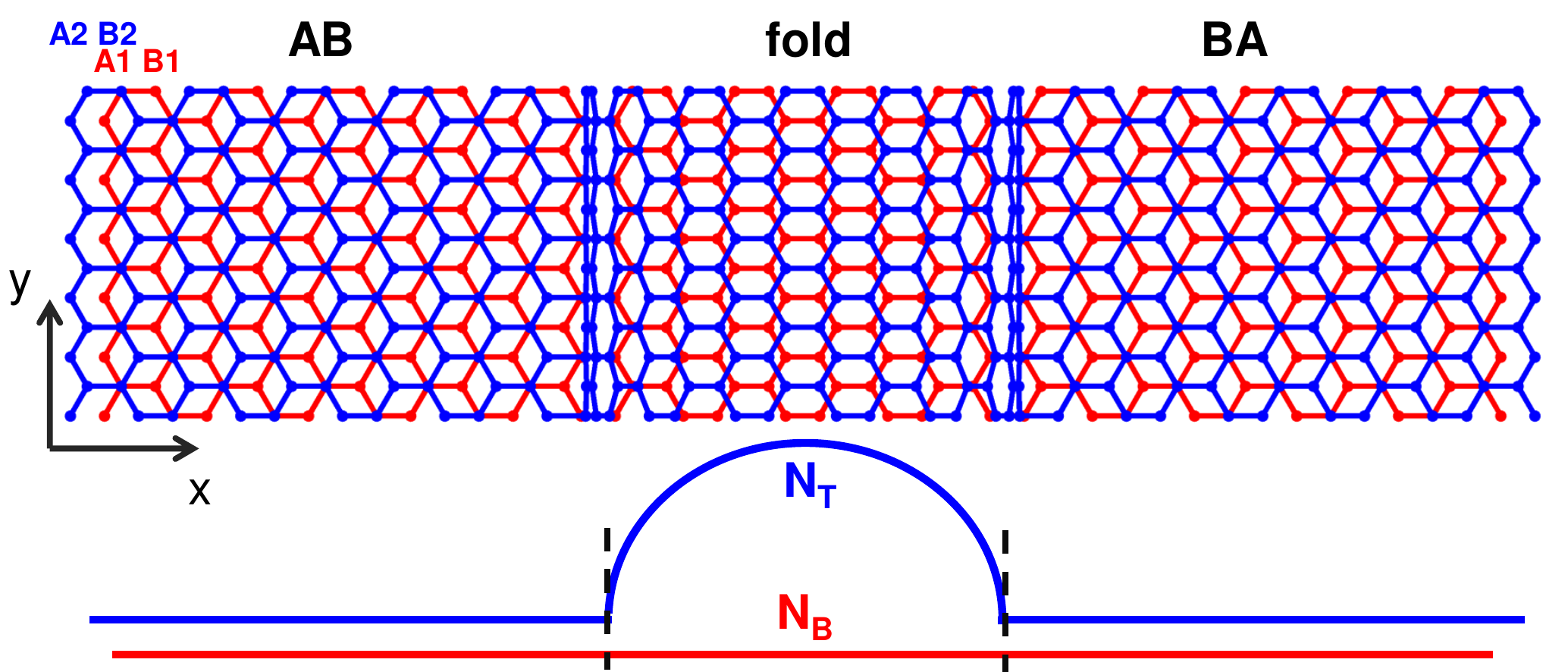}
\caption{\label{fig:system}
(Color online). Top and side view of bilayer graphene with an AB/BA stacking boundary due to a corrugation of the upper layer.  The stacking change from AB to BA is clearly visible in the top view. The side view below shows the geometry chosen for the corrugation or fold, namely, half a nanotube. $N_B$ and $N_T$ are the lengths given in the number of unit cells of the bottom and top layer at the fold.}
 \end{figure}
In this paper we consider stacking boundaries in bilayer graphene produced by a fold or corrugation of the top layer.
Such defect can be modeled as half a nanotube protruding from the plane of the upper layer (see Fig. \ref{fig:system}), seamlessly joined to the upper semiinfinite graphene planes.

In order to elucidate its electronic and transport properties, we first study the one-dimensional (1D) case, i.e., a fold in a metallic armchair bilayer nanoribbon. Then, we describe  AB/BA stacking boundaries in the two-dimensional (2D) case, namely, a corrugation in bilayer graphene. Interestingly, we  observe a drop of the conductance through the corrugation in the un-gated system. We attribute this conductance gap to the symmetry under the simultaneous exchange of layers and sublattices. By applying a gate voltage to the bottom layer, energy gaps open and a series of localized states appear in the gap. The states are mostly localized around the fold, although they extend appreciably into the leads, decaying with an oscillatory behavior. In many cases the absolute maximum of the local density of states (LDOS) takes place outside the corrugation. We distinguish between valley-polarized topological states, originating from the stacking change, and eigenstates due to the finite size of the corrugation. We find that the states localized at the corrugation do not contribute to the conductance across the boundary, but constitute perfect conductance channels along this fold, in agreement with recent experimental measurements \cite{Ju_2015}.


\section{\label{sec:geomod} Geometry and model}

\subsection{\label{sec:geo} Geometry}
In order to fix the notation, we refer to AB stacking when the A site from the bottom graphene layer, A1, lies exactly below the B2 site of the top layer. Likewise, BA stacking corresponds to the B1 site lying below the A2 (see Fig. \ref{fig:system}).  The connection between AB and BA-stacked regions needs a deformation of the lattice. Here we consider the stacking boundary to consist of a corrugated region, as it is shown in Fig. \ref{fig:system}. The central part connecting the two perfect AB and BA-stacked regions consists of two graphene layers, one of them longer, creating a fold. Because of the graphene lattice symmetry, this kind of stacking boundary can be created in a simple way by connecting layers in AB and BA systems at a zigzag line. Thus, the fold can be easily modeled as a portion of an armchair nanotube; the bottom layer below the fold is a piece of flat zigzag ribbon seamlessly connected to the bottom layer of the leads. If we assume that the bilayer is situated on a substrate, the most natural configuration would have a flat bottom layer. The sizes of the bottom and top layers in the stacking boundary (the central part), $N_B$ and $N_T$, are given in terms of the translational unit cell in the $x$ direction. For a given $N_B$ value we choose a larger $N_T$ length which allows for the change of stacking. As mentioned above, this extra length should naturally accommodate as a fold.

\subsection{\label{sec:mod} Model and method}

We use a one-orbital tight-binding (TB) Hamiltonian to describe the system. We assume uncoupled layers in the fold. Only the nearest-neighbor hopping parameter $\gamma_{0}=-2.66$ eV is considered, and set the on-site energy $\epsilon_0=0$. The bilayer leads with Bernal stacking are described by $H_{L,R}=H_{1}+H_{2}+H_{12}$, where $H_{1}$ and $H_{2}$ are the single-layer Hamiltonians corresponding to the bottom and top layer and the interlayer Hamiltonian $H_{12}$ connects only those atoms which are on top of each other. The interlayer hopping parameter is taken as $\gamma_{1}=0.1\gamma_{0}$. We consider that the gate voltage $V$ is applied only to the bottom layer, as it would be in a reasonable experimental setup. 

Since the system with a single corrugation has no translation symmetry in the $x$ direction (perpendicular to the fold, see Fig. \ref{fig:system} ), we compute the LDOS and the conductance using the Green function matching method. The total Hamiltonian of the system can be written as \cite{Datta, Chico_1996, Jaskolski_2005}:
\begin{equation}
H=H_{L}+H_{R}+H_{C}+V_{LC}+V_{RC},
\end{equation}
where $H_{L}$ and $H_{R}$ are the Hamiltonians of the left (L) and right (R) leads, respectively. $H_{C}$ is the Hamiltonian of the conductor in central part of the system. In our case it is the region of the corrugation: the fold composed by half a nanotube plus a flat piece of graphene below, $V_{LC}$ and $V_{RC}$ are the connections of the central part to the left and right lead, respectively. The Green function of the central $C$ part is a function of the energy $E$:
\begin{equation}
\mathcal{G}_{C}(E)=(E-H_{C}-\Sigma_{L}-\Sigma_{R})^{-1}
\end{equation}
where $\Sigma_{L}=V_{LC}g_{L}V_{LC}^{\dagger}$ and $\Sigma_{R}=V_{RC}g_{R}V_{RC}^{\dagger}$ are the self-energies of the leads with $g_{L,R}$ being the Green functions of the leads.

We calculate the conductance $G$ using the Landauer-B\"uttiker formalism,
\begin{equation}
G(E)=\frac{2e^{2}}{h}T(E)=\frac{2e^{2}}{h}Tr[\Gamma_{L}\mathcal{G}_{C}\Gamma_{R}\mathcal{G}_{C}],
\end{equation}
where $T(E)$ is the transmission function from the left lead to the right one and $\Gamma_{L,R}=i[\Sigma_{L,R}-\Sigma_{L,R}^{\dagger}]$ describe the couplings between the conductor and the $L$, $R$ leads.

To study the spatial distribution of the states and their energy dependence, we draw on the complex band structure of the semiinfinite bilayer graphene leads \cite{Heine_1963, Lee_1981_I, Lee_1981_II}, where the complex values of the $k$ wavevectors can be related to the decay lengths of such states. 
It is a tool successfully used for the analysis of the decay lengths of localized states in bulk systems 
\cite{Schulman_1981, Chang_1982, Wortman_2002}, linear periodic molecules \cite{Tomfohr_2002}, as well as in graphene \cite{Park_2013}; recently it has also been applied to topological insulators \cite{Fu_2007, Dang_2014}. Here we additionally employ it to explain the oscillations of the corrugation-confined states.


\section{\label{sec:ribbons}Corrugated bilayer graphene nanoribbons}

From both, the computational and the conceptual viewpoint, it is easier to address in the first place the properties of the 1D system, that is, a bilayer graphene nanoribbon with a wrinkle or corrugation in the top layer. Such ribbon can be considered as a strip of 2D bilayer graphene with a corrugation. Since we chose the zigzag direction to be parallel to the corrugation, this implies that the finite-size ribbon cut perpendicularly to this fold should be of the armchair type, as show in Fig. \ref{fig:system}. As we are interested in the transport properties, we select the widths so that the bilayer nanoribbon leads are metallic at zero gate voltage. The width $W$ of the ribbon is given in terms of the translational unit cell in the $y$ direction. The ribbons constituting the bilayer are assumed to have minimal armchair edges and a perfect vertical stacking, as the one depicted in Fig. \ref{fig:system}. This is the so-called $\alpha$-stacking in the literature \cite{Sahu_2008,Gonzalez_2010,Santos_2012}.  

\subsection{\label{sec:cond_gap}Conductance drop around $E_F$ at zero bias}

\begin{figure}[thpb]
      \centering
\includegraphics[width=\columnwidth]{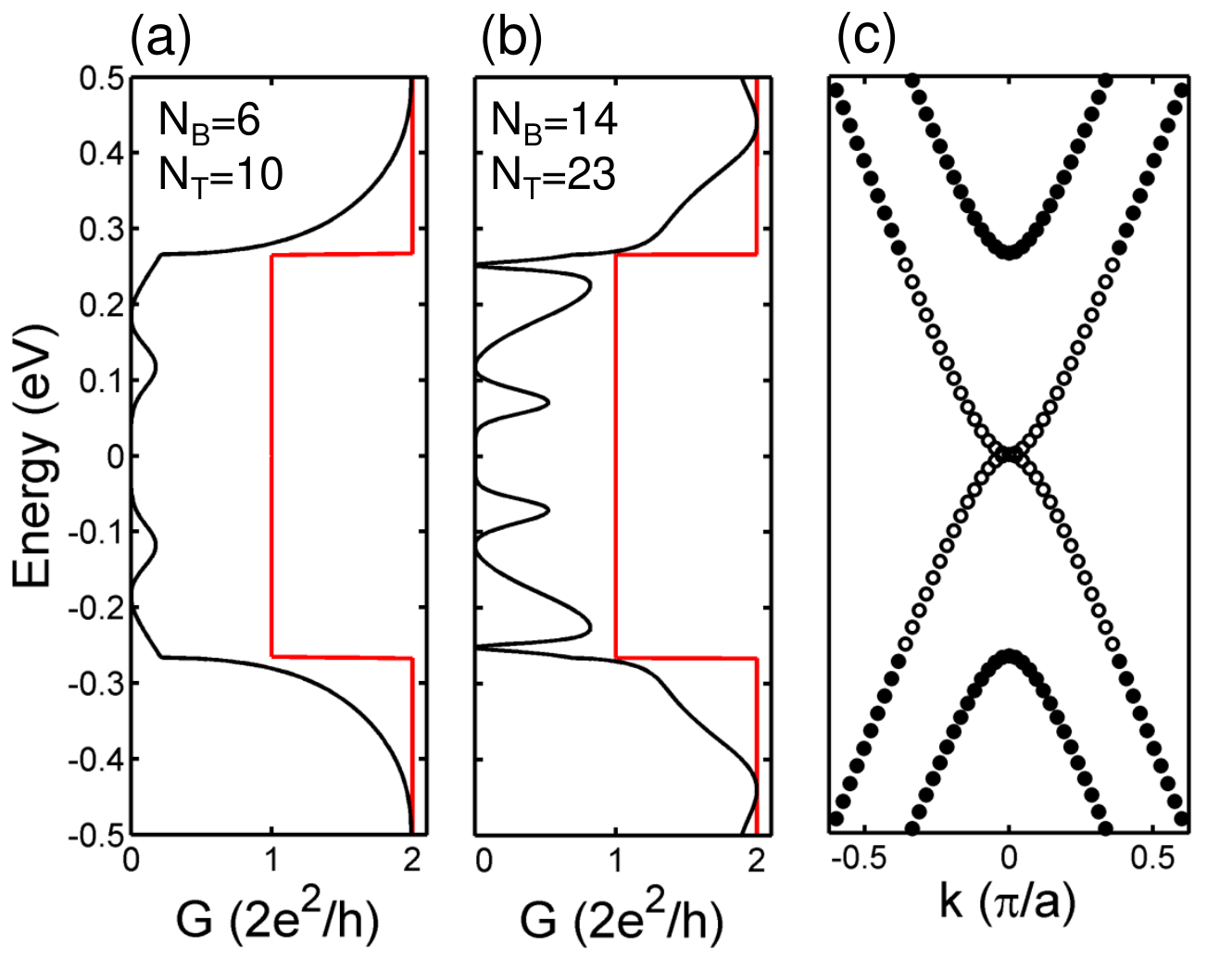}
\caption{\label{fig:cond_drop}
(Color online). Conductance of two bilayer ribbons with the width $W=4$ and two different fold lengths: (a) $N_B=6, N_T=10$, (b) $N_B=14, N_T=23$. The red line shows the conductance of the infinite bilayer ribbons acting as leads. Panel (c) shows the band structure close to the Fermi level of a bilayer ribbon with Bernal stacking. Open circles indicate that the corresponding states are localized on disconnected sublattices, i.e.,  B1 and A2 sites for AB stacking and A1 and B2 sites in the BA case.
}
   \end{figure}

In Fig. \ref{fig:cond_drop} we show the results for a bilayer ribbon of width $W=4$ with a fold. We consider two different  corrugations, $N_B=6, N_T=10$ in panel (a) and $N_B=14, N_T=23$ in panel (b), in the absence of gate voltage. In the energy range $|E|<\gamma_1$ we observe a conductance drop, more dramatic in the case of  the smaller corrugation. In fact, the maximum value of the conductance and the number of maxima in $|E|<\gamma_1$ increases with the difference between the fold layer lengths $|N_T-N_B|$.

The conductance drop appears although both leads as well as the central part are metallic. To understand this, we should consider the spatial distribution of the scattering states, i.e., those in the left and right leads. Obviously, both leads, bilayer nanoribbons with AB and BA stacking, have exactly the same band structure, shown in Fig. \ref{fig:cond_drop} (c). However, recall that their spatial distribution in the energy range $|E|<\gamma_1$ is in the uncoupled nodes, these being B1 and A2 sites in the AB stacking and A1 and B2 sites in the BA case  \cite{McCann_2006, Mucha_2010}. We indicate this spatial distribution with open circles in the band structure  (Fig. \ref{fig:cond_drop} (c)). The localization in opposite sublattices is the reason behind the conductance drop. Indeed, the longer the corrugation length, the larger the conductance in this energy region, as discussed above. This points to a symmetry related to the simultaneous exchange of sublattices and layers, broken in the presence of a fold. 
 
We have calculated an ideal abrupt AB/BA boundary, without any extra atoms for the fold, just by connecting directly an AB and a BA bilayer nanoribbon. If the hoppings across the boundary are kept equal, then the conductance gap is perfect, due to the existence of the aforementioned symmetry: such idealized boundary is invariant under an exchange of sublattices and layer position. If the hoppings at the boundary are modified, the symmetry is broken and a small nonzero conductance is observed in the gap. Likewise, adding more atoms in a folded layer or changing the hoppings in a strained planar boundary makes the conductance increase due to the larger symmetry  breaking. 

\subsection{\label{sec:loc_states}Gap-localized states in biased folded ribbons}

\begin{figure}[thpb]
      \centering
\includegraphics[width=\columnwidth]{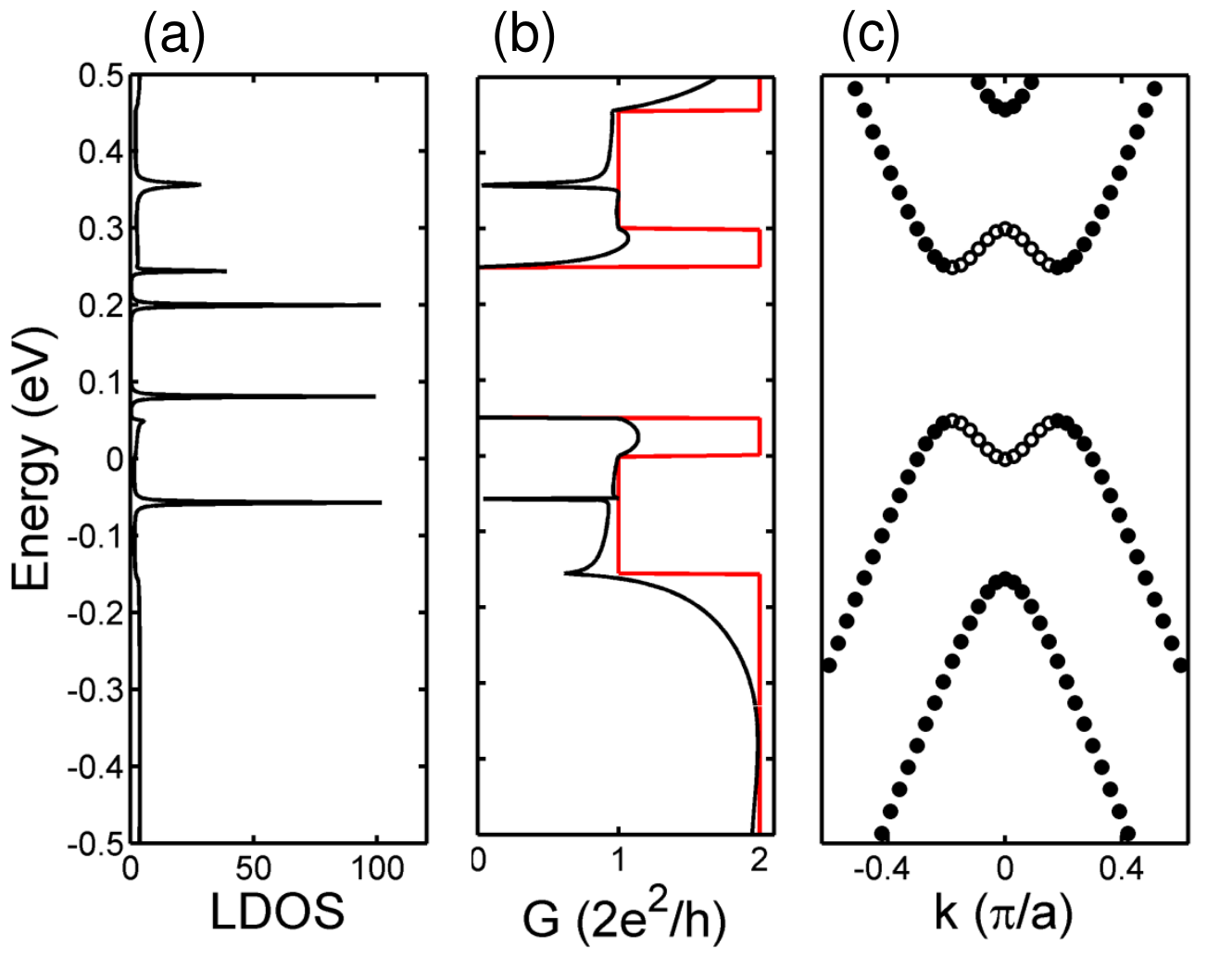}
\caption{\label{fig:loc_states}
(Color online). (a) LDOS and (b) conductance of a bilayer corrugated ribbon with a $V=0.3$ V gate voltage applied to the bottom layer. 
The width of the ribbon is $W=4$ and the folded region lengths are $N_B=6$ and $N_T=10$. The red line in (b) shows the conductance of the infinite bilayer leads with the same bias. (c) Band structure of the infinite bilayer ribbons which constitute the leads. Open circles indicate that the corresponding states are localized on disconnected sublattices, whereas filled circles indicate states with weight in any sublattice.
}
   \end{figure}

We consider a gate voltage $V$ uniformly applied to the entire bottom layer, i.e., to both leads and the central flat part lying below the fold (Fig. \ref{fig:loc_states}). With respect to the leads, it is well-known that the voltage opens a gap and changes the shape of the bands to a ``Mexican hat"-type dispersion \cite{Castro_2010, Ohta_2006}. Fig. \ref{fig:loc_states} (c) shows the band structure of the bilayer leads with $V=0.3$ V, with a gap in the energy range $0.05<E<0.25$ eV. For small $k$ values the valence and conduction band states are still localized on uncoupled sites (indicated in the Figure with open circles), similarly to the $V=0$ case. However, due to the ``Mexican hat" dispersion, there are two channels available at low energies, and the second one is extended to the coupled sublattices as well, indicated by filled circles in Fig. \ref{fig:loc_states} (c). 
   
Figs. \ref{fig:loc_states} (a) and (b) show the LDOS and conductance of a biased ($V=0.3$ V) bilayer ribbon of width $W=4$ and corrugation with $N_B=6$ and $N_T=10$. The LDOS is summed over all the nodes of the central part which constitutes the corrugation. We observe a series of LDOS peaks; some of them appear in the nonzero conductance region and give rise to antiresonant conductance drops whereas in the conductance gap ranging from 0.05 eV to 0.25 eV we find two peaks at energies $E_1=0.081$ eV and $E_2=0.2$ eV. The number of localized states in the gap is equal to two only for small $N_B$ and $N_T$. Increasing the length of the fold leads to the appearance of more localized states. These gap peaks correspond to the quantization of the topological boundary states which appear in the 2D corrugated bilayer graphene. The origin and properties of the localized states can be analyzed more easily resorting to the 2D case (see Section \ref{sec:plane}).

\subsection{\label{sec:oscill} Spatial distribution of the localized states}

\begin{figure}[thpb]
      \centering
\includegraphics[width=\columnwidth]{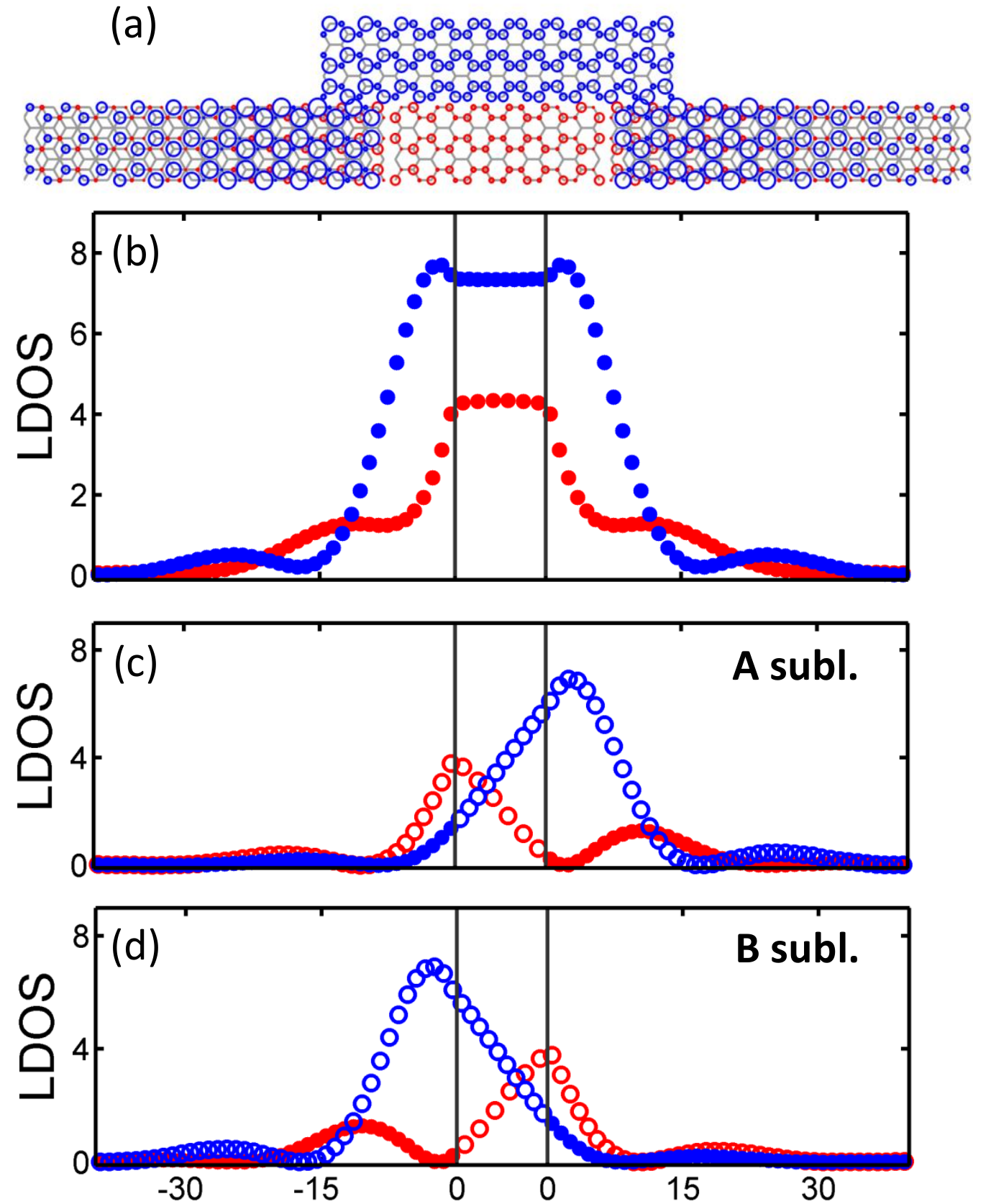}
\caption{\label{fig:ldos_oscil}
(Color online). Spatial distribution of the localized state with energy $E_1=0.081$ eV found in the corrugated  bilayer ribbon of width $W=4$ and corrugation given by  $N_B=6$ and $N_T=10$. (a) Atom-resolved LDOS; the circles plotted on each node have radii proportional to the corresponding LDOS values. Red and blue colors indicate that the node belongs to the bottom and top layers respectively. (b) Unit-cell averaged LDOS at the bottom (red) and the top layer (blue); the zero labels mark the limits of the stacking boundary.  (c,d): Unit-cell-averaged LDOS at both layers (red - bottom; blue - top), plotted separately for (c) A and (d) B sublattices; in these two latter panels, open circles indicate that the LDOS is located at unconnected nodes.
}
   \end{figure}

The two states appearing in the gap that we described in the previous Subsection are produced by the change of stacking across the boundary and should be localized therein. 
To corroborate this, we analyze their spatial distribution. We focus on one of the localized states, that with $E_1=0.081$ eV found in the corrugated $W=4$ ribbon with a folded region of $N_B=6$ and $N_T=10$. Figure \ref{fig:ldos_oscil} (a) shows the LDOS distribution in the fold and in the adjacent regions of the leads; the radii of the circles are proportional to the LDOS. The color indicates the layer where the LDOS is evaluated (bottom - red, top - blue). In order to elucidate the distribution in layers and sublattices, we present in Fig. \ref{fig:ldos_oscil} (b) the LDOS at the bottom (red dots) and top (blue dots) layers, and in Figs. \ref{fig:ldos_oscil}  (c) and (d) we show how the state is distributed in the A and B sublattices, respectively. 

Being a stacking boundary state, it is mostly localized at the central part which constitutes the corrugation. We find that the maximum of the LDOS is not always located in the corrugation region, but instead, it may have its maximum value in the adjacent cells to the stacking boundary, extending appreciably into the leads. Recall that for this energy range, bulk lead states are located in the uncoupled sublattices of the bilayer, which are reversed in AB and BA stacking. We verify that indeed, this localized state is also mostly located in the uncoupled sublattices far from the fold. As they are reversed at opposite sides of the boundary, they must exchange from one to the other in the corrugation region. Such lattice swapping is most evident in Figs. \ref{fig:ldos_oscil}  (c) and (d). 

The wavefunction of the localized state decays moving away from the corrugation, but with an oscillatory behavior. This oscillation, visible in the LDOS separated by layers, is even clearer when 
plotted for different sublattices. The oscillations in the top and bottom layers of the leads are in antiphase, but obviously with the same oscillation period and decay constant. These can be obtained from the complex band structure of the leads \cite{Heine_1963}. The idea is to allow the wavevectors $k$ to have complex values. These analytic extensions stem from the extrema of the real band structure and 
the corresponding wavevectors describe the decay and oscillations of the interface or boundary states lying in the energy gap. It can be considered as solving the Hamiltonian eigenvalue problem for a fixed energy inside the gap, so the solutions yield the complex wavevectors. In fact they can be obtained more easily from the transfer matrix formalism \cite{Lee_1981_I, Lee_1981_II}.  

For the state shown in Fig. \ref{fig:ldos_oscil} we can directly extract from the LDOS behavior the spatial frequency of the oscillations, $\omega=0.044$ and the decay constant $\alpha=0.125$, assuming that the LDOS $ \propto \cos^2 (\omega x) \, e^{-\alpha x}$. From the complex band structure, we get 
$\omega= {\rm Re}(k)/\pi=0.044$ and $\alpha = 2 \,{\rm Im}(k)=0.118$, which compare rather well to the fitted values. The unit length in the previous quantities is the BLG nanoribbon unit cell size. As these complex wavevectors are the analytical continuation from the gap at the K points, with nonzero values, they have a real component which gives rise to the oscillations. 

Notice that, since the bilayer leads are the same in any AB/BA boundary, irrespectively of the particular geometry of the transition region, we can expect the same oscillating behavior in 
other types of stacking boundaries, such as those produced by strained bilayer regions \cite{Vaezi_2013}. 

We have verified that the numerical agreement between the fitted values and those extracted from the complex band structure is excellent in other states with sharper oscillations produced by higher values of $V$.

The second gap-localized state with $E_2=0.2$ eV is also distributed mainly at the corrugation. However, it is more strongly localized at the uncoupled nodes of the bottom layer than $E_1$. Its behavior is approximately as that of $E_1$ exchanging the top and bottom layers and the A and B sublattices simultaneously.

\section{\label{sec:plane}Corrugated bilayer graphene}

 We consider now the 2D bilayer graphene with either one isolated stacking boundary, or with periodically repeated folds. The orientation of the fold is as described in Sec. \ref{sec:geo}, namely, along the zigzag direction. Thus, the flat bilayer and the corrugation have translational symmetry in the $y$ direction.  
  
In 2D systems, we can consider two different geometries to evaluate the conductance: either perpendicular to the stacking boundary or along it. We start with the same configuration as for the nanoribbons, that is, transport perpendicular to the corrugation. We will next examine a periodically repeated fold in order to elucidate the conductance along the corrugations. 
   
\subsection{Single fold in bilayer graphene} 
  
We first consider a 2D graphene bilayer with one corrugation. Like in the finite ribbons, we take the current to flow perpendicularly to the fold. In this sense, the flat bilayers act as leads. 

\begin{figure}[thpb]
      \centering
\includegraphics[width=\columnwidth]{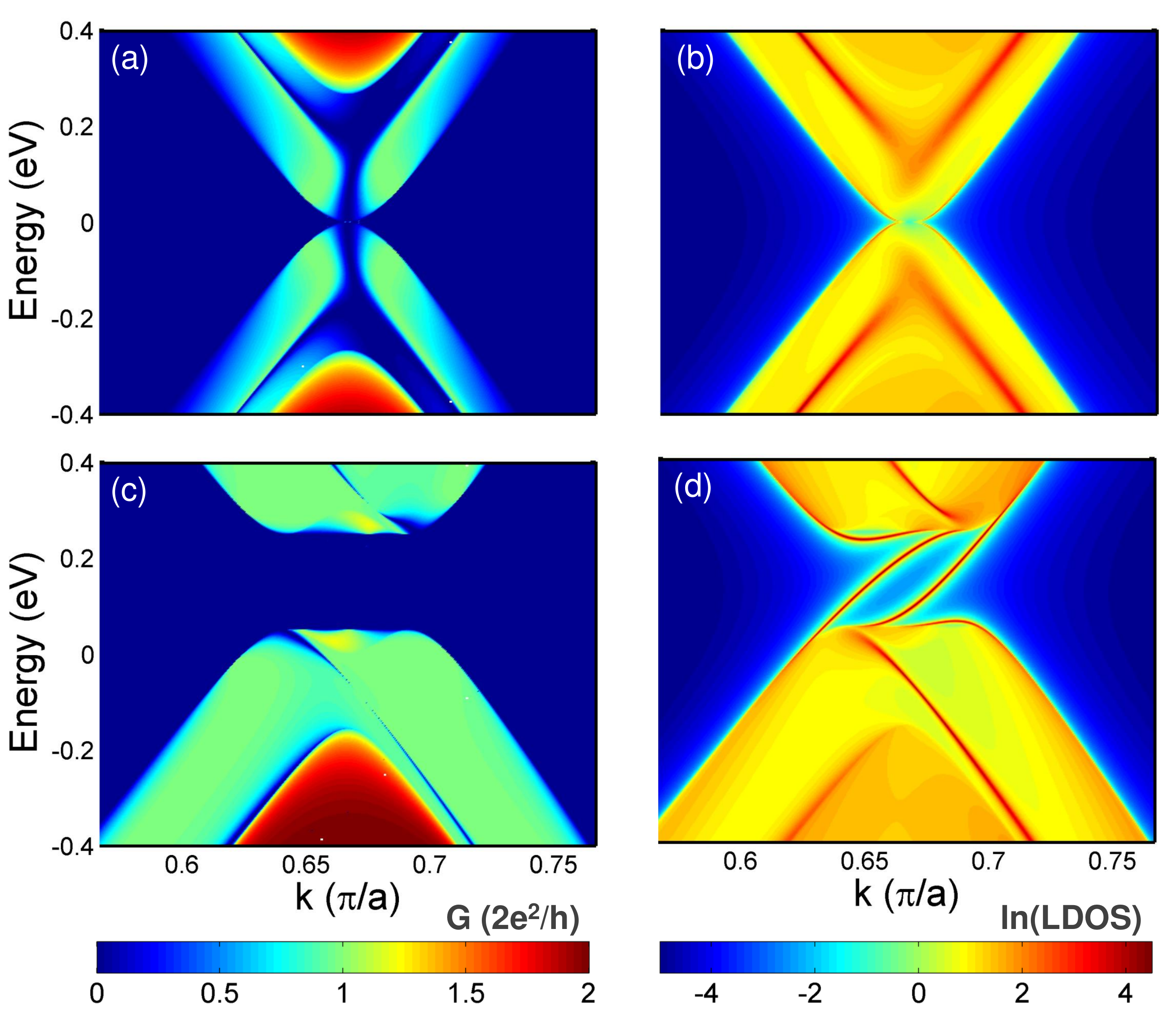}
\caption{\label{fig:2D}
Conductance between leads (left panels) and LDOS at the corrugation (right panels) for the $N_B=6$, $N_T=10$ case as functions of the energy $E$ and wavevector $k$;  (a) and (b) are without gate voltage;
(c) and (d) with $V=0.3$V applied to the bottom layer.
Notice that the LDOS scale is logarithmic.
}
   \end{figure}

Figure \ref{fig:2D} shows the conductance between the bilayer leads and projected LDOS \cite{Santos_2009} at the corrugation as functions of $k$ and $E$, with and without gate voltage. 
The corrugation lengths are $N_B=6$ and $N_T=10$, and the LDOS$(E,k)$ is summed over all nodes in the unit cell of the fold. The results are shown for $k\geq0$. As we consider the current flowing from the left to the right lead, we focus on the Dirac cone at positive $k$ stemming from K'  of the reciprocal graphene lattice.  Negative $k$ values are therefore related to K, which is its mirror reflection in the present geometry. The only difference is thus the sign of the carriers' velocity. Comparing the plots on Figs. \ref{fig:2D} (a) and (c), with $V=0$ and $V=0.3$V respectively, we observe the expected  energy shift of the conductance and LDOS as well as the appearance of the conductance gap due to the voltage. This gap opening can be also seen in the corresponding LDOS$(E,k)$  plot (Fig. \ref{fig:2D} (d)). More importantly, we have two bands crossing the gap and connecting the two bulk continua when $V$ is applied, which constitute two localized states in the corrugation of topological origin due to the change from AB to BA stacking. These states are valley-polarized; they give no contribution to the conductance for this current direction, but will carry a valley-polarized current along the fold.  

\begin{figure}[thpb]
      \centering
\includegraphics[width=\columnwidth]{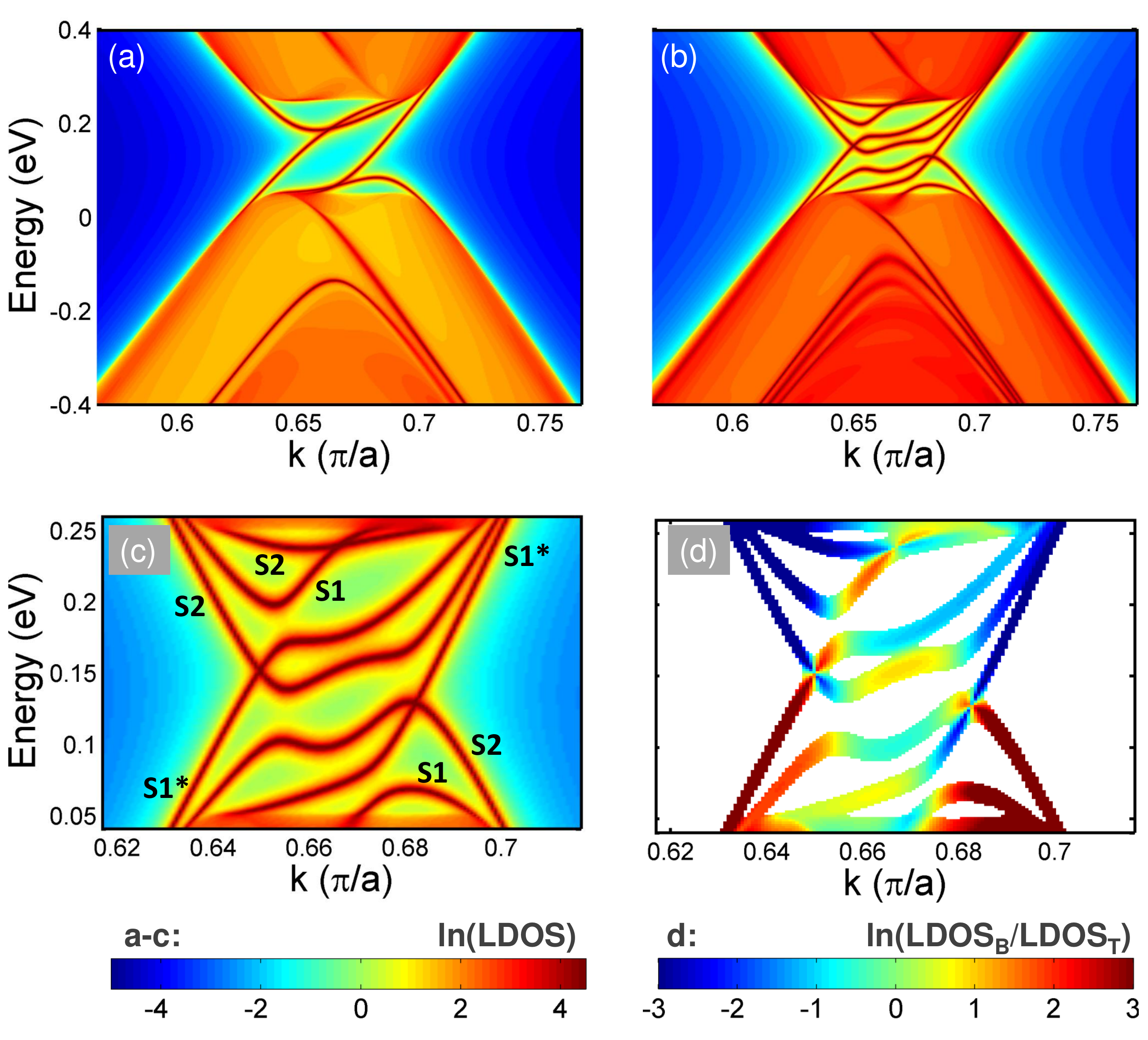}
\caption{\label{fig:2D_lcorr}
LDOS as a function of the energy $E$ and wavevector $k$ at the corrugation, summed over all its nodes, for two large instances:  (a) $N_B=14$, $N_T=23$ and (b) $N_B=30$, $N_T=48$. Because of the large values of the LDOS at the localized states, the LDOS scale is logarithmic.
(c) Zoom of the gap states presented in (b), with two types of symmetries marked as S1 and S2 (topologically protected states market with the asterisk). (d)  Layer-resolved LDOS of the same system,  
 $\ln$(LDOS$_B$/LDOS$_T$), plotted only when LDOS$_B$+LDOS$_T$ is larger that a threshold value $10^{-3}$.
}
   \end{figure}

Figure \ref{fig:2D_lcorr} shows the LDOS for two instances of large corrugations in bilayer graphene, $N_B=14$, $N_T=23$ (panel (a)) and $N_B=30$, $N_T=48$ (panel (b)). With increasing $N_B$ and $N_T$, we observe the appearance of more bands in the gap. Differently to the stacking boundary bands of topological origin, which cross the gap, they start and end in the same cone (upper or lower). 
These bands correspond to the eigenstates of the two finite layers composing the corrugation region. They are quantum-well-like states arising from the finite-size effect imposed by the change from the bilayer regions to two uncoupled monolayers of different size.

Figure \ref{fig:2D_lcorr} (c) is a  zoom of the LDOS shown in Fig. \ref{fig:2D_lcorr} (b), for a corrugation given by $N_B=30$ and $N_T=48$. The successive crossings and anticrossings of these localized bands indicate the existence of two types of symmetry, labeled as S1 and S2. The topological states have the same type of symmetry (S1 with asterisk), while the quantum-well-like states appear with S2 and S1 alternately. In order to elucidate these symmetries, we have calculated the wavefunctions of a related periodic system - a BLG with a periodic corrugation of the same characteristics. We find that S1 states have horizontal (along the $x$ direction) nodal surfaces in the corrugation, while S2 states have also vertical nodal surfaces.

Figure \ref{fig:2D_lcorr} (d) shows the same bands from panel (c), now distinguishing between bottom (red) and top (blue) layer localization. There is a remarkable exchange of localization between layers at the anticrossings, due to the hybridization between these bands. 

\subsubsection*{\label{sec:bound_cond}Folded ribbons from discretizing 2D corrugated graphene}

The analysis of bilayer graphene with a single fold allows us to understand the size dependence of the gap states appearing in the 1D ribbons described in the previous Section. In the same way that an armchair ribbon may be treated as a strip cut of the infinite plane, we expect that the properties of the corrugated ribbon with a particular width can be obtained by imposing the proper quantization rules along the $y$ direction to the 2D LDOS($E,k$) and $G(E,k)$ calculated for the corrugated bilayer graphene (Fig. \ref{fig:2D}).
If a bilayer armchair ribbon of width $W$ is cut from a graphene bilayer oriented as in Fig. \ref{fig:system}, the corresponding wavefunctions should vanish for $y=0$ and $y=(W+\frac{1}{2})a$, where $a$ is the graphene lattice constant. That gives us the quantization rule for the $k_y$ vector, 
$k_q=\frac{q\pi}{(W+\frac{1}{2})a},$ where $q=1,...,W$. According to this, for a particular $W$ value, the  LDOS of the ribbon is the sum of LDOS for all $k_q$ calculated for all the allowed $q$ values. The same can be said for the conductances. We can compare Fig. \ref{fig:cond_drop} with Fig. \ref{fig:2D}. For $V=0$ we observe the conductance drop when one of the allowed $k_q$ values is equal to $\frac{2}{3}\frac{\pi}{a}$; it would be a cut of the 2D plot  in Fig. \ref{fig:2D} (a) through the middle of the cone, yielding a low conductance as in Fig. \ref{fig:cond_drop} (a). After applying this rule to the nonzero $V$ case (see Fig. \ref{fig:loc_states} (a) with Fig. \ref{fig:2D} (d)), we see the $W$ dependence of the localized states. Note that allowed $k$ values include $\frac{2}{3}\frac{\pi}{a}$  when $W=3m+1$, which is always the case for the metallic ribbons considered in this work. For narrow ribbons, the energies of the gap-localized states for fixed $N_B$ and $N_T$  do not depend on $W$ because they stem from $\frac{2}{3}\frac{\pi}{a}$. However, for $W\geq25$, there are other $k_q$ values which cross the stacking boundary bands lying in the gap, giving rise to more localized states whose energies do depend on $W$.

\subsection{\label{sec:cond}Periodic corrugations in bilayer graphene: conducting topological states}

\begin{figure}[thpb]
      \centering
\includegraphics[width=\columnwidth]{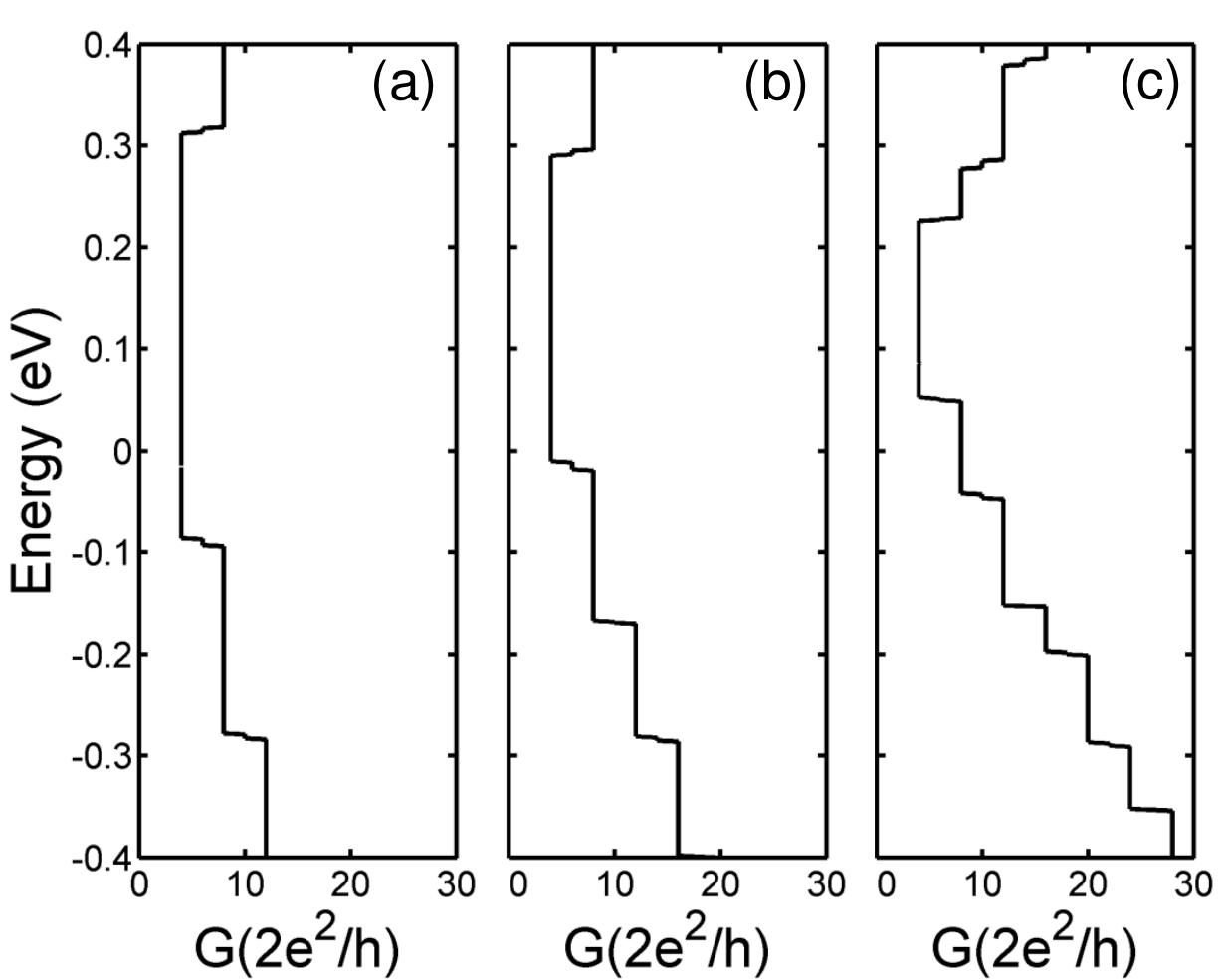}
\caption{\label{fig:periodic_cond}
Conductance of periodically corrugated graphene bilayers with gate voltage $V=0.3$ V with periodic corrugations. (a) $N_B=6$ and $N_T=10$,  separated with $d=5$ unit cells; (b) $N_B=6$ and $N_T=10$,  separated with $d=10$ unit cells;
 (c) $N_B=14$ and $N_T=23$ separated with $d=10$. 
}
   \end{figure}

We have already seen how the conductance through the corrugation is lowered due to the stacking change, and explained this reduction. We have also proven that gap states give no contribution to the current perpendicular to the fold. However, we expect these states to be conducting along the corrugation. In order to verify this, we consider a bilayer graphene with periodic corrugations separated with long Bernal stacking regions (AB and BA alternately) of size $d$ in translational BLG unit cells along the armchair direction $x$. We calculate the conductance in the $y$ direction. For this system and transport setup, the conductance is a function of the wavevector in the $x$ direction, $k_x$. As we are interested in the current flowing along the corrugation, we take $k_x=0$. This assumption implies that the current is measured with a local probe that makes the contribution of other scattering directions negligible. In the energy range around Fermi level we expect four conductance quanta ($G_0=2e^2/h$), as we have two corrugations in the unit cell, each having two topologically protected states. Fig. \ref{fig:periodic_cond} shows the conductance for a corrugated bilayer graphene plane with  $V=0.3$ V 
with two different periodic corrugations. Close to the Fermi level we always have four conductance quanta, which are valley-polarized.  Increasing the size of the corrugation and the distance between them,  there is a general rise of the conductance related to the increasing number of bands. However, around the Fermi level the conductance is always equal to four $G_0$. This is due to the fact that the topologically protected states are the only ones crossing $E_F$. In a transport experiment, it is possible to employ a probe that contacts only one corrugation; in such a case, the conductance close to $E_F$ in a gated system would be at most two conductance quanta belonging to a single valley.

\section{\label{sec:summary} Summary}

We have explored the electronic and transport properties of bilayer graphene and the corresponding metallic graphene nanoribbons with an AB/BA stacking boundary composed of a corrugation. We have shown that the transition between AB and BA zigzag-ended stackings can take place a fold formed by half a nanotube and a nanoribbon. 

Without an external gate voltage, these systems are gapless, but present a conductance gap due to the symmetry mismatch related to the simultaneous exchange of sublattices and layers. The application of a gate voltage produces the appearance of gap states at the corrugation which are topologically protected in the absence of intervalley mixing. 

For larger folds, more localized states appear at the corrugation, that we relate to quantum-size effects. These states are gapped, contrarily to those stemming form the change of stacking. We have analyzed the spatial dependence and transport properties of all these corrugation-localized states, verifying their oscillatory decay far from the corrugation, that we have related to the complex band structure of the bilayer regions. 

Finally, we have shown that these states are conductive along the folds, constituting robust conductance channels that can be easily identified by their geometry, being extended folds or corrugations across bilayer graphene regions.


\begin{acknowledgements}
This work was partially supported by the Polish Ministry of Science and Higher Education (The "Mobility Plus" Program) and the Spanish Ministry of Economy and Competitiveness under Grant No. FIS2012-33521 and FIS2013-48286-C2-1-P. We thank to J. Iribas and L. Brey for illuminating discussions.

\end{acknowledgements}
%


\end{document}